\documentclass[mathserif,compress]{article}
\usepackage{psfrag,color}
\usepackage{epsfig} 
\usepackage{graphicx}
\usepackage{times}
\usepackage{cmap}
\usepackage[TS1,T2A]{fontenc}
\usepackage[koi8-r]{inputenc}
\usepackage{natbib}
\usepackage[english]{babel}
\bibpunct{[}{]}{;}{a}{}{,}  

\newcommand{\pc}{\ensuremath{\rm pc}}

\newcommand{\kms}{\mbox{km~s}^{-1}}
\newcommand{\arcsec}[1]{^{\prime\prime}\!\!\!#1\,}   

\renewcommand{\mag}[1]{^{\rm m}\!\!\!#1\,}

\newcommand{\href}[1]{\url{#1}}

\begin{document}
\title{Modeling Hydrogen-Rich Wolf-Rayet Stars in M33} 
\author{Olga Maryeva\\[2mm]
\begin{tabular}{l}
 Stavropol State University, Stavropol 355001, Russia, {\em olga.maryeva@gmail.com}\\
\end{tabular}
}
\date{}
\maketitle

\begin{abstract}    
  We  present the results of a spectral variability study of two very 
  luminous stars in the M33 galaxy -- LBV V532 and late WN star
  (possibly, dormant LBV) FSZ35. 
  We studied spectral variability of
  V532, derived its atmosphere parameters and showed that the bolometric
  luminosity varies between the two states by a factor of 
  $\sim$1.5. Using the non-LTE radiative transfer code CMFGEN, we
  determined wind parameters for both objects. Since both stars are
  located at distances of about 100 pc from the nearest association,
  we supposed that they may be massive runaway stars with velocities 
  of the order $100\kms$.
\end{abstract}  
 
\section{Introduction}

            Luminous blue variables (LBVs) are rare objects of very
            high luminosity ($\sim10^6 L_{\odot}$) and mass loss
            rates  ($10^{-5}{M}\div10^{-4} {M}_\odot\mbox{yr}^{-1}$), exhibiting strong
            irregular photometric and spectral variability ~\citep{Conti,HumphreysDavidson}.
            LBVs span quite a large range of magnitudes and
            variability types~\citep{HumphreysDavidson}.  
            Despite much research, there is still no consensus 
            about the evolution of massive stars.
            In particular, there is no answer to the question, whether all 
            WN stars passed through the LBV stage?   
            Contemporary evolutionary scenarios say that massive 
            O-stars with mass  90$M_{\odot}$ cross to  WNL without 
            going through the LBV phase \citep{maeder}. 
$$
M>90M_{\odot}:~\mbox{\footnotesize O}\to \mbox{\footnotesize Of} \to 
\mbox{\footnotesize WNL} \to \mbox{\footnotesize(WNE)} \to\mbox{\footnotesize WCL} \to 
\mbox{\footnotesize WCE} \to \mbox{\footnotesize SN(hypernova~at~low~Z?)}
$$
whereas if mass is $40\div90~M_{\odot}$ then 
$$
60-90M_{\odot}:~O\to \mbox{\footnotesize Of/WNL} \Leftrightarrow \mbox{\footnotesize LBV} \to \mbox{\footnotesize WNL(H-poor)} 
\to \mbox{\footnotesize (WNE)} \to \mbox{\footnotesize WCL-E} \to\mbox{\footnotesize SN(SNIIn?)} $$
\vspace*{-0.5cm}
$$
40-60M_{\odot}:~\mbox{\footnotesize O} \to \mbox{\footnotesize BSG}\to 
\mbox{\footnotesize LBV}\Leftrightarrow \mbox{\footnotesize WNL}\to\mbox{\footnotesize (WNE)}\to~ 
\mbox{\footnotesize  WCL{\footnotesize-}E}\to \mbox{\footnotesize  SN(SNIb)~or}\to\mbox{\footnotesize   WCL{\footnotesize-}E}\to\mbox{\footnotesize   WO~SN(SNIc)}$$

            Generally, LBV are  believed 
            to be a relatively short evolutionary stage in the 
            life of massive stars.  
            However, recent investigations of the light curves of a few 
            supernovae (SN2006gy, 2006tf, SN2005gl)  
            indicate that their progenitors underwent LBV-like eruptions 
           ~\citep{SmithMcCray,smith2008,GalYam2009}. 
            These observations support the view that at least some luminous LBV
            stars are the end point of the evolution  but not a transition phase. 

Therefore, a continuous spectral and photometric monitoring and  
numerical modeling of the atmospheres of these objects   
are very important for understanding the evolutionary     
relationship WR and LBV stars, as well as  of the physical 
causes of LBVs variability.  
            Observations of Galactic LBV stars are inevitably connected with 
            difficulties in determination of the distance and interstellar extinction, 
            which results in huge uncertainties in their bolometric luminosities. 
           Thus, studying of these rare objects in nearby galaxies is 
            potentially more prospective, though extragalactic objects are more 
            distant and hence are more difficult targets for spectral observations. 
           
        FSZ35 and V532 (known as Romano's star) are stars with similar spectra and, 
        probably, with similar evolution, but with 
        different histories of research. 
        V532 star, known by its variability, is the famous object in M33 galaxy. 
        Photometric observations have been carried out 
        from 1961 year~\citep{romano,ZGphoto}, 
        spectral observations -- from 1992~\citep{szeifert}. 
       ~\citet{polcaro} classify V532 as a LBV.  
        The spectrum of object changes from a B emission line 
        supergiant in the optical maximum~\citep{szeifert}, 
        through Ofpe/WN (WN10,WN11) and WN9 toward a WN8-like 
        spectrum in deep minima~\citep{me}. 
       ~\citet{Sholukhova} note similarity between 
        the spectra of Romano's star  and  FSZ35.
        The object  FSZ35 is little-studied WNL star in M33. 
       ~\citet{Ivanov} were the first to obtain photometric 
        data for this object (IFM-B 174, in their notation). 
        The star is listed in the H$\alpha$ emission-line object
        catalogue~\citep{fsz} as number 35. In article ~\citet{massey98} 
        the star is listed as object E1 and is classified 
        as a Wolf-Rayet star of nitrogen sequence, WN8 subtype. 

\begin{figure}[h] 
\begin{center}
\includegraphics[width=0.8\textwidth]{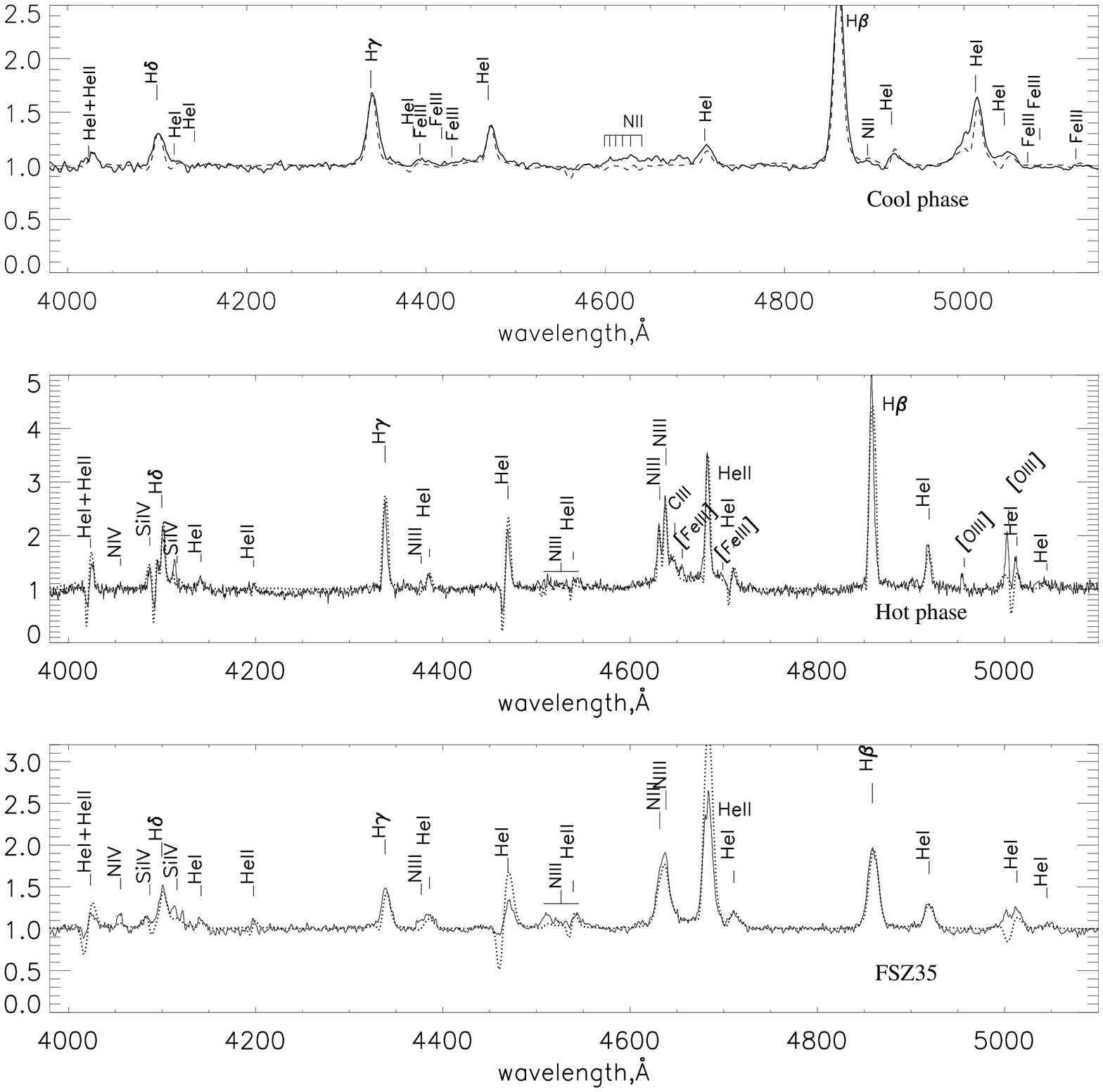}
\end{center}
\caption{The normalized optical spectra (solid line)
      compared with the best-fit CMFGEN models (dashed line).
      Top panel: the spectrum obtained 
      in February 2006 when V532 is $17\mag{.}27$ in V band.
       Middle: the spectrum obtained in October 2007, when
       V-band magnitude is ${\it V}=18\mag{\ .}68$. 
       Bottom: The spectrum of FSZ35 with the final model.} 
\label{fig:coldmodel1}
\end{figure}
\begin{table*}[h]\centering
\caption{Observational log for the spectral data used in the
  work.   S/N is signal-to-noise ratio per resolution element.} 
\label{tab:obstabscor}
\bigskip
\begin{tabular}{lcccccccc}
\hline
          &          &Exposure        &       & Spectral      & $\delta \lambda$&          &   Seeing                & Spectral  \\ 
 ~~~~~Date&          &  time, [s]     & Grism & range, [\AA]  & ~[\AA]          &   S/N    &  [$\arcsec{\,}$~]       & standard star \\
\hline
\multicolumn{9}{c}{\bf V532}  \\ 
06.02.2005  &   {\footnotesize SCORPIO}    & $2\times 300$   &  {\footnotesize VPHG550G}& 3700-7200   &     10             &   8  & 1.7               & {\footnotesize G248}       \\
08.01.2008  &   {\footnotesize SCORPIO}    & $2\times 900$   & {\footnotesize VPHG1200R}&  5700-7500  & 5                  &  20  & 2.1               &  {\footnotesize BD25d4655} \\
08.10.2007  &   {\footnotesize FOCAS}      & $1200$          & {\footnotesize  VPHG450} &  3750-5250  & 1.7                &  30  & 0.53              &  {\footnotesize BD40d4032} \\
\multicolumn{9}{c}{\bf FSZ35}    \\
4. 10. 2007  &   {\footnotesize SCORPIO}   &  $900+1200$     &{\footnotesize  VPHG1200G}&  4000-5700  & 5.5                & 19   & 1.2               &  {\footnotesize G191B2B}   \\
\hline
\end{tabular}
\end{table*}
       In this work we investigate the optical spectra of V532 
        and FSZ35 using the non-local thermodynamic equilibrium 
        (non-LTE) radiative transfer code CMFGEN \citep{Hillier5}.
        We pay special attention to comparison between 
        these two objects. 

             This paper is organized as follows. The observational data
             and data reduction process are described in the next section. 
             In Section~\ref{sec:v532} and Section~\ref{sec:fsz35} 
             we present and analyse the modeling
             results for V532 and FSZ35. In Section~\ref{sec:disc} we discuss the results.

\section{Observations and Data Reduction}

              In this work, we use spectral data were taken from the archive of 6-m BTA 
              telescope of Special Astrophysical Observatory (SAO) of Russian
              Academy of Sciences (RAS) ({http://www.sao.ru}) and 
              from the SMOKA science archive~\citep{smoka} of SUBARU 
              telescope ({http://smoka.nao.ac.jp}).
              The 6-m telescope data were obtained with the SCORPIO multi-mode 
              focal reducer in the long-slit mode~\citep{scorpio}.
              One exposure V532 was obtained with the SUBARU telescope  
              with the Faint Object Camera (FOCAS)~\citep{focas} in October 2007. 
              Observational log information on the data used in 
              this work is summarized in Table~\ref{tab:obstabscor}.

               All the  SCORPIO spectra were reduced using the {\tt ScoRe} 
               package is written in IDL language. The package includes all 
               the standard stages of long-slit data reduction process. 
               FOCAS data were reduced in IDL development  environment 
               using procedures similar to those consisting {\tt ScoRe } 
               but taking into account the specific features of FOCAS. 
               We describe the observational data of V532 and FSZ35 
               in more detail in~\citet{me} and in~\citet{fsz35}, respectively.

\section{Romano's star}\label{sec:v532}

           We investigate the optical spectra of Romano's star in
           two different states,
           the brightness minimum of 2008 (${\it B}=18.5 \pm 0.05$mag) and a moderate
           brightening in 2005 (${\it B}=17.1\pm 0.03$mag). 
           We refer to both states as hot and cool phases, respectively.

            In Figure~\ref{fig:coldmodel1} we show the observed 
            spectra of V532 at different phases and the best-fit model spectra. 
            Stellar parameters derived for both hot- and cool-phase
            models are given in Table~\ref{tab:parmodel}. For comparison, 
            the values of these parameters for some other stars
            are given in the table. Table~\ref{tab:parmodel} shows that V532 in the minimum of 
         brightness is similar to a classical WN8 star, but the wind
         velocity is lower, characteristic rather for a WN9 star. 
         We see that relative hydrogen abundance (H/He) for V532 is 
         similar to that of WN8h stars. In February 2005, during 
         the outburst, parameters of the star correspond to the spectral class WN11. 
         The model spectrum is similar to the spectrum of P~Cyg in 1998.
         V532 shows a WN11 spectrum in the maximum, while the
         classical LBVs like AG~Car and P~Cyg become WN11 only in deep
         minima and in the long-lasting quiet state, respectively. 
         Note however that V532 had a strong maximum in 1993 ($0\mag{\ .}9$ brighter than
         in February 2005) and exhibited a B-supergiant spectrum.  
         The two phases, hot and cool, are mainly distinguished by the
         photosphere radius. In the hot phase the radius is about three times larger than
         in the cool phase.
         Three basic parameters vary simultaneously and make measurable
         contributions to the observed inflation of the star. 
         For the two states, $\dot{M}$ differ by a factor 2.4, and the wind
         velocity is 1.8 times larger for the hot state. Our models favour correlation of  hydrostatic radius with
         mass loss rate.
\begin{table*}\centering
\caption{Derived properties of V532 in the maximum and the minimum of
  brightness, and comparison with related stars in M33, LMC and Milky
  Way (MW) galaxies, including the LBVs P~Cyg and AG~Car in visual minima (Dec
  1990). 
H/He denotes hydrogen number fraction relative to helium.
} 
\label{tab:parmodel}
\bigskip
\begin{tabular}{lcccccccccccc}
\hline
Star      & Gal. & Sp.   & $T_*$ &   $R_*$ & $T_{eff}$ &$R_{2/3}$   & $\log L_*$&$\log \dot{M}_{cl}$          & f  & $v_{\infty}$& H/He & Ref       \\
         &      & type  & [kK]  & [$\rm R_{\odot}$] &   [kK]  &[$\rm R_{\odot}$]&[$\rm L_{\odot}$]&[$\rm M_{\odot}\,yr^{-1}$]&    & [$\kms$]   & &     \\
\hline
 WR124    & MW   &  WN8h & 32.7  &    18.0   &           &             &   5.53    &     -4.7         & 0.1&  710        & 0.7  & \citep{Crowther99} \\ 
 WR40     & MW   &  WN8h & 45.0  &    10.6   &           &             &   5.61    &     -4.5         & 0.1&  840        & 0.75 &  \citep{HHH}\\ 
 WR16     & MW   &  WN8h & 41.7  &    12.3   &           &             &   5.68    &     -4.8         & 0.1&  650        & 1.2  &  \citep{HHH}\\ 
 AG~Car   &  MW  & WN11  & 24.64 &    67.4   &  21.5     &   88.5      &   6.17    &     -4.82        & 0.1&  300        &  2.3    & \citep{groh} \\
Dec 1990  &      &       &       &           &           &             &           &                  &    &             &      &         \\ 
P Cyg     &  MW  &B1I$a^+$&      &           &  18.7     &   76.0      &   5.8     &    -4.63         & 0.5&  185        & 2.5  &  \citep{NajarroPCyg} \\
 FSZ35    &  M33 &  WN8h & 36.5  &    19     & 35        &  20         &   5.76    &    -4.58         & 0.1&  800        &  0.8  &     \\
 V532     &  M33 &  WN8h & 34.0  &  20.8     &  31.7     &  23.9       &   5.7     &    -4.72         & 0.1&  $360^*$    &  1.9 &     \\ 
hot-phase &      &       &       &           &           &             &           &                  &    &             &      &     \\
 V532     &  M33 &  WN11h& 22.0  &  59.6     &  20.4     &  69.1       &  5.89     &    -4.4          & 0.5&  200        &  1.4 &     \\
cool-phase&      &       &       &           &           &             &           &                  &    &             &      &      \\ 
\hline
\multicolumn{13}{l}{$^{*}$ -- $v_{\infty}$ was estimated using HeI lines~\citep{me}}
\end{tabular}
\end{table*}
          
             The luminosity of V532 in 2005 
            ($L_*=7.7\cdot10^{5}L_{\odot}$) is  1.5 times higher. 
            Therefore, V532 should be considered one more LBV (after
            the objects mentioned by
           ~\citet{koen04,drissen,Clark09}) that changes its luminosity
            during (even moderate amplitude) eruption. 
            In this sense, V532 behaves similarly to AG~Car that has bolometric
            luminosity variations during its S~Dor cycle~\citep{groh}. 

\section{FSZ35}\label{sec:fsz35}
            We analyzed a spectrum of FSZ35 in the $4000\div 5500$\AA\ 
            wavelength range and identified about 40 spectral lines.  
            The spectral appearance of FSZ35 shows strong similarities with V532. 
            Therefore we classify FSZ35 as a WN8 as
            well. Besides this, the NIV 
            $\lambda 4057$ line clearly seen in our
            spectrum is never present in WN9 spectra, thus excluding
            FSZ35 identification as a later-subclass object. 

       The best-fit parameters of the CMFGEN model are given in Table~\ref{tab:parmodel}.
       Derived parameters of FSZ35 
       atmosphere correspond to a typical WN8 star. Because mass 
       fraction of hydrogen is 16.5\% (H/He=0.8) in  the spectrum of FSZ35 
       we classify this object as H-rich WN8 star. 

\begin{figure}
\begin{center}
\includegraphics[width=0.6\textwidth]{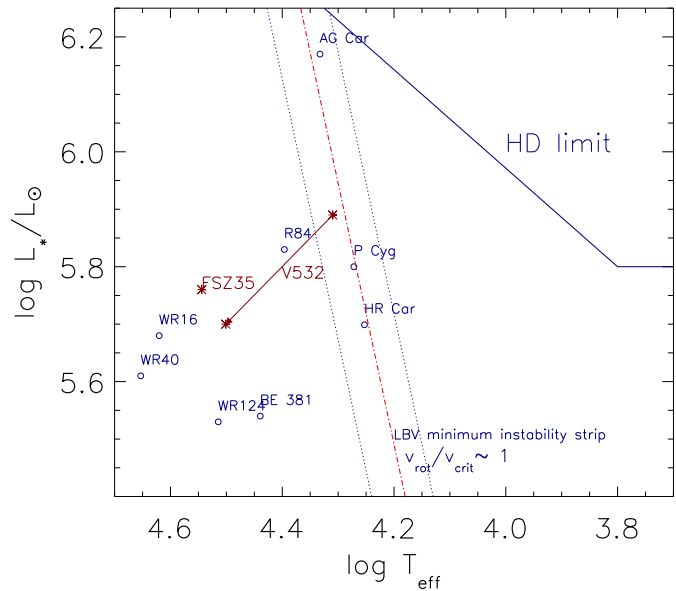}
\end{center}
\caption{HR diagram showing the positions of V532 and FSZ35, 
according to the stellar parameters determined in this work. 
 Position for the LBV minimum instability strip is provided (red dashed line).
The location of the Humphreys-Davidson limit \citep{HumphreysDavidson} is shown (solid line).
WN9 (BE381, R84), WN8 (WR124, WR16, WR40) and LBV (AG~Car, HR~Car, P~Cyg) stars are shown for comparison. Data on these  
                               objects were taken from
                            \citep{CrowtherLMC}, \citep{CrowtherLMC}, \citep{Crowther99}, 
                            \citep{HHH}, \citep{HHH}, \citep{groh}, \citep{grohHR}, \citep{NajarroPCyg},consequently
}
\label{fig:HRdiagram}
\end{figure}

\section{Discussion}\label{sec:disc}
   Figure \ref{fig:HRdiagram} presents the positions of FSZ35 and V532  in different phase in 
the Hertzsprung-Russell (HR) diagram. \citet{grohHR} suggest that the LBV minimum instability strip is characterized by 
$\log (L_*/L_{\odot})=4.54\cdot\log(T_{eff})-13.61$. 
\citet{grohHR} suggest that the LBV
minimum instability strip corresponds to the region where 
critical rotation is reached for LBVs with strong S-Dor-type variability. 
When LBVs are evolving toward maximum, the star
moves far from the LBV minimum instability strip (to the
right in the HR diagram), and $v_{rot}/v_{crit}$ decreases considerably. 
The region in the HR diagram on the left side of the
LBV minimum instability strip would be populated by unstable
LBV stars with $v_{rot}/v_{crit}>1$, which would make it a ``forbidden region'' for LBVs. 
Indeed, no confirmed strong-variable LBV is seen in this region (see \citep{grohHR} and references therein). 
V532 in maximum of brightness ({\it V}=$17\mag{\,}$, Feb.~2005) 
lies on the  LBV minimum instability strip. And it moves to ``forbidden region'' in the minimum of brightness. 
Unlike V532, FSZ35 does not demonstrate any prominent spectral variability. The lack of photometrical variability
             may be ascribed to the more advanced evolutionary status of
             FSZ35.

According to the modern theories, there is no mechanism of formation of single stars are born the groups consisting of two and more components, entering into larger base units (complexes, clusters, etc.). 
             FSZ35 is located  at the distance of about
             $35\arcsec{\,}$\ ($\sim115$ pc) from the 128
             association. 
             Suppose that once FSZ35 was a member of 128 association and 
             was ejected via slingshot-type dynamical interaction. 
             If its peculiar velocity is  $\sim 100\kms$, as for WR124~\citep{wr124vel,marchenko}, 
             it could have been expelled from the parent cluster about
             a million years ago.

             Offset positions with respect to the probable parent associations
             (at distances $\sim 100\pc$) and unexpectedly large peculiar
             velocities (of the order $\sim 100\kms$) seem to be common for
             very luminous and massive stars like V532, FSZ35 and Galactic late WN
             stars like WR20a and WR124. A scenario was proposed by~\citet{GG11}  
             that applies three-body dynamical interaction in the
             cores of young massive
             clusters and star-forming regions (similar to 30~Doradus) to reproduce the
             observed population of massive runaways. This scenario has the 
             disadvantage of relying heavily on the formation of massive stars in
             the cores of massive young clusters that are, in contrast
             with LMC, absent in M33, where massive stars are rather formed
             in dispersed stellar associations. 
             Instead we would rather propose that very massive stars are
             formed in dense groups containing several stars each. This is
             confirmed, for example, by the increasing multiplicity
             with stellar mass~\citep{Zinnecker} simultaneously in young star
             clusters and in associations. Reasonably 
             the higher fraction of massive binaries the higher accompanied fraction of massive multiple
             systems. 
%
              The characteristic peculiar velocities ($\sim 100\kms$) of these
             ``childhood runaways'' may be reproduced if the initial
             spatial sizes of the systems are $\sim 10^{14}\rm cm$.
\vspace*{-0.27cm}
\section*{Acknowledgments}
        I would like to thank John D. Hillier for his great code CMFGEN, which is so 
        comprehensive and user-friendly, that we applied to fit and analyse
        the data very comfortably.  
        Based in part on data collected at Subaru Telescope, which is operated
        by the National Astronomical Observatory of Japan, and taken
        from the SMOKA, operated by the Astronomy Data Center,
        National Astronomical Observatory of Japan. We also use the
        data from the archive of the Special Astrophysical Observatory. 
        This work was supported by the Russian Foundation for Basic Research (project no. 11-02-00319-а). 
        Author is grateful to Russian Astronomical Society for financial support of participation in JENAM.

\end{document}